# Magnetic structure of diamond $EuTi_2Al_{20}$

S. W. Lovesey[1, 2, 3] and D. D. Khalyavin[1]

[1] ISIS Facility, STFC, Didcot, Oxfordshire OX11 0QX, UK

[2] Diamond Light Source, Harwell Science and Innovation Campus, Didcot, Oxfordshire OX11 0DE, UK

[3] Department of Physics, Oxford University, Oxford OX1 3PU, UK

**Abstract**. We study magnetic symmetries of $EuTi_2Al_{20}$ that appear in the ordering of Eu ions below a temperature ≈ 3.3 K, according to an analysis of substantial experimental evidence [M. Kawamata *et al*., J. Phys. Soc. Jpn. **95**, 024701 (2026)]. They descend from the cubic diamond structure that depends explicitly on mixing of orbitals with different spatial parities, e.g., diamond s- and p-atomic states. The lattice (2, 2, 2) Bragg spot, for example, is lattice forbidden in the absence of parity mixing. Our work is motivated by a first proposal that low-temperature magnetic symmetries of diamond $EuTi_2Al_{20}$ are orthorhombic and belong to magnetic crystal classes that include time reversal. Body-centre anti-translation in diffraction patterns is a corollary. Current experimental data does not distinguish between two orthorhombic candidates, even though one is centrosymmetric and the other one is polar. We demonstrate that magnetic parity-odd entities (multipoles) formed by mixing d-and f-atomic states at europium positions, with discreet symmetries that match those of a Dirac monopole, are different for the two candidates. Moreover, the multipoles are visible in resonant x-ray diffraction patterns. The selection rule from parity mixing in the parent cubic diamond structure is preserved, and permitted Bragg spots are different for the two candidate magnetic symmetries. Our symmetry informed diffraction patterns include rotation of the crystal about the reflection vector (an azimuthal angle scan) and a full account of x-ray polarization to spur more revealing experiments.

## I. INTRODUCTION

Europium ions in $EuTi_2Al_{20}$ form a network that is of interest in studies of magnetic frustration. Enigmatic properties of the compound include an enhanced magnetoresistance and an unconventional Hall effect, with the possible involvement of skyrmions [1,2]. The room temperature structure of $EuTi_2Al_{20}$ is the cubic diamond structure $Fd\bar{3}m$ (No. 227). Magnetic long-range order is detected at a temperature below ≈ 3.3 K. According to an analysis of solid experimental evidence, the low temperature magnetic symmetry belongs to a magnetic crystal class that includes time reversal $1'$ [3]. Diamond ions are $sp^3$ hybridized meaning electrons are concentrated in bond regions between carbon ions rather than being spherically distributed around the nucleus. The absence of spherical symmetry produces forbidden lattice reflections epitomized by (2, 2, 2) intensity reported by Bragg [4,5]. Europium ions in the proposed orthorhombic magnetic symmetries occupy asymmetric Wyckoff positions that permit mixing of d- and f-atomic states [6,7]. It is responsible for polar magnetic entities, Dirac multipoles [8], that diffract x-rays tuned in energy to an Eu atomic resonance.

The concept of hidden magnetic order is firmly established, and it usually refers to magnetic properties beyond axial dipoles not encountered with standard laboratory experimental probes. As often as not, hidden order from higher-order magnetic entities, e.g., octupoles, can be exposed by scattering techniques using suitably prepared beams of electrons, neutrons or photons [9]. Solid evidence for this is inferred from Bragg diffraction patterns gathered from neutron powder and resonant x-ray scattering on single crystals [3]. Kawamata *et al.* [3] propose magnetic symmetries for $EuTi_2Al_{20}$ associated with a propagation vector $k_1$ = (1, 0, 0). Candidate symmetries for $EuTi_2Al_{20}$ include body-centre anti-translation, and magnetic Bragg reflections indexed by integer Miller indices (*h*, *k*, *l*) are permitted for odd (*h* + *k* + *l*). In most cases of value, rotation of the x-ray polarization depicted in Fig. 1 is required for this diffraction technique to respond to magnetic $EuTi_2Al_{20}$. The candidate symmetries imply big reductions in symmetry of the compound, namely, low-temperature centrosymmetric No. 52 or polar No. 34 orthorhombic symmetries down from room-temperature cubic hexoctahedral diamond No. 227.

The big reduction in symmetry of $EuTi_2Al_{20}$ contrasts with properties of the classic example of antiferromagnetic manganese fluoride ($MnF_2$). Rutile manganese fluoride has magnetic tetragonal symmetry that is a subset of the parent structure, and its magnetic crystal class does not include $1'$ [10]. Lattice forbidden reflections in the cubic diamond structure, e.g., (2, 2, 2), have been exploited in a direct observation of anapoles ("without poles" in Greek) by magnetic neutron diffraction [8,11]. The reflection (9, 1, 0) included in the study of $EuTi_2Al_{20}$ satisfies the same extinction rule, Fig. 3(h) in Ref. [3]. Dirac multipoles (anapole is an alternative name for a Dirac dipole) are compulsory entities in $EuTi_2Al_{20}$ diffraction patterns, because resonant Eu ions use asymmetric Wyckoff positions in the candidate magnetic symmetries. Our studies show contributions to diffraction patterns for the two candidates are not the same [3], which remains to be confirmed by experiments. Returning to known differences between $MnF_2$ and $EuTi_2Al_{20}$, manganese fluoride possesses an x-ray chiral signature that is not allowed in the orthorhombic candidates for $EuTi_2Al_{20}$ [3,10].

In more detail, the two candidates for $EuTi_2Al_{20}$ are orthorhombic magnetic symmetry $P_I$nna or $P_I$nn2 [3,12]. Diffraction patterns for parity-even (axial) Eu absorption events in our studies must be identical, of course. $P_I$nna is centrosymmetric and $P_I$nn2 is polar, and different parity-odd (Dirac) responses are predicted. In keeping with observations [3], an unrotated diffraction amplitude for reflection vectors (0, 0, *l*) is zero. A difference in calculated and observed Bragg spots indexed by (*h*, *k*, 0) is fundamental. Symmetry protects null diffraction enhanced by parity-even absorption events with odd *h*, which appears to conflict with experimental data reported for the reflection vector (9, 1, 0) [3]. Notably, axial and Dirac amplitudes are predicted to interfere, because their phases are common. Dirac amplitudes are not considered in published data analysis [3]. We study two classes of reflections labelled Domain A (0, 0, *l*), and Domain C (*h*, *k*, 0) that are associate with propagation vectors $k_1$ = (1, 0, 0) and $k_3$ = (0, 0, 1) [3]. Orthorhombic reflection vectors are found in an Appendix. Domain B (*h*, *k*, *l*) associated with $k_2$ = (0, 1, 0) is omitted, because the diffraction pattern is unduly complicated while offering no additional insights. Miller indices for the orthorhombic structures violate conditions for lattice reflections in the cubic diamond structure Fd$\bar{3}$m.

## II. RESONANT X-RAY BRAGG DIFFRACTION PATTERNS

A resonant atomic process may dominate all other contributions to the x-ray scattering length should the photon energy E match an atomic resonance energy $\Delta$ [13,14,15]. Assuming virtual intermediate states in the process are spherically symmetric, to a good approximation, the x-ray scattering length $\approx \{(\mu\eta)/(E - \Delta + i\,\Gamma/2)\}$in the region of the resonance, where $\Gamma$ is the total width of the resonance [16]. The numerator $(\mu\eta)$ is an amplitude for Bragg diffraction in the scattering channel with primary (secondary) polarization $\eta(\mu)$. In a generally accepted convention, $\sigma$ labels polarization normal to the plane of scattering, and $\pi$ labels polarization within the plane of scattering, as in Fig. 1, which includes the Bragg angle $\theta$.

Allowance is made for an azimuthal angle scan of Bragg spot intensities in our symmetry informed calculations, in which the illuminated crystal is rotated through an angle $\psi$ about the reflection vector [13-16]. Kawamata *et al.*, do not provide relevant measurements on this standard mode of investigation. Instead, the primary x-ray beam is $\pi$-polarized and an analyser collects the $\sigma$-polarized signal [3]. The two extreme settings of the analyser measure Bragg spot intensities $|(\pi'\pi)|^2$ and $|(\sigma'\pi)|^2$. There is currently no experimental data for the unrotated channel of polarization $\sigma \rightarrow \sigma'$ for which we report several predictions.

Central to symmetry informed studies of diffraction is an electronic structure factor,

$$\Psi^K_Q = \sum \exp(i\, \boldsymbol{\kappa} \cdot \mathbf{d})\, \langle O^K_Q \rangle_{\mathbf{d}}. \qquad (1)$$

It delineates Bragg diffraction patterns for magnetic neutron and resonant x-ray scattering. The sum is over positions **d** used by Eu ions in a unit cell, and the reflection vector $\boldsymbol{\kappa}$. The electronic structure factor includes information about the relevant Wyckoff positions found in the Bilbao table MWYCKPOS [12]. Wyckoff positions are related by operations listed in the table MGENPOS [12]. Taken together, the two tables provide all information required to complete the electronic structure factor Eq. (1) and our symmetry informed studies. The generic atomic multipole $\langle O^K_Q \rangle$ in Eq. (1) is of integer rank K with (2K + 1) projections $-K \le Q \le K$ [16, 17, 18]. Angular brackets denote a time-average of an enclosed spherical quantum operator, and therefore our multipoles belong to the magnetic ground state. Signatures of discrete symmetries are time $\sigma_\theta = \pm 1$ and parity $\sigma_\pi = \pm 1$. A complex conjugate $\langle O^K_Q \rangle^* = [(-1)^Q \langle O^K_{-Q} \rangle]$. The constraint $[\sigma_\theta (-1)^K] = +1$ is imposed by parity-even electric dipole - electric dipole (E1-E1) and electric quadrupole - electric quadrupole (E2-E2) absorption events, and associated multipoles are $\langle T^K_Q \rangle$ [16]. The time signature $\sigma_\theta = -1$ for Dirac multipoles denoted $\langle G^K_Q \rangle$ with $\sigma_\pi\, \sigma_\theta = +1$ (E1-E2; electric dipole - magnetic dipole). By way of an detailed illustration, axial and Dirac multipoles for a realistic model of Cu in monoclinic CuO are reported in Ref. [17]. Atomic resonances for rare earth ions are $L_{2,3}$ edges 2p $\rightarrow$ 5d, E1; 2p $\rightarrow$ 4f, E2; $M_{4,5}$ edges 3d $\rightarrow$ 5f, E1 [6,7]. Dipole sum-rules for L edges include $[\langle \mathbf{T}^1 \rangle_{L2} + \langle \mathbf{T}^1 \rangle_{L3}] \propto \langle \mathbf{L} \rangle_d$, and $[\langle \mathbf{G}^1 \rangle_{L2} + \langle \mathbf{G}^1 \rangle_{L3}] \propto \langle \mathbf{R} \times (\mathbf{L} + 2\mathbf{S}) \rangle$ for an E1-M1 event with a magnetic dipole $M1 = (\mathbf{L} + 2\mathbf{S})$ and position vector **R** [17].

**A. Parent structure** Fd$\bar{3}$m (No. 227). Europium ions take Wyckoff positions 8a (symmetry $\bar{4}$3m) in the cubic diamond structure that are acentric. We find,

$$\Psi^K{}_Q(8a) = \langle O^K{}_Q \rangle \exp(i\pi\{H_o + K_o + L_o\}/4)\, [1 + (-1)^{Ko + Lo} + (-1)^{Ho + Lo} + (-1)^{Ho + Ko}]$$
$$\times [1 + \sigma_\pi \exp(-i\pi\{H_o + K_o + L_o\}/2)], \qquad (2)$$

with integer Miller indices ($H_o$, $K_o$, $L_o$). When $\sigma_\pi = -1$ Bragg spots satisfy a reflection condition $\{H_o + K_o + L_o\} = 2(2n + 1)$, and we have mentioned $n = 1$. All $EuTi_2Al_{20}$ reflections are magnetic, with Miller indices that violate lattice symmetry which results in $\Psi^K{}_Q(8a) = 0$.

**B. Magnetic symmetry P$_I$nna** (No. 52.320). The symmetry is orthorhombic, centrosymmetric, and it belongs to the magnetic crystal class mmm1′. Europium ions use Wyckoff positions 4e, with general coordinates (1/4, 0, 7/8) [3]. Basis vectors with respect to the parent structure are (½($b + c$), ½($b - c$), $-a$) [3]. These orthogonal vectors are used to define local axes ($\xi$, $\eta$, $\zeta$) for electronic multipoles. Local ($\xi$, $\eta$, $\zeta$) coincide with (x, y, z) in Fig. 1 in the nominal setting of the sample with respect to the diffractometer, and the azimuthal angle $\psi = 0$. Our calculated electronic structure factor is,

$$\Psi^K{}_Q(P_I nna) = \exp\{i\pi(2h - l)/4\}\, [1 + \sigma_\theta (-1)^{h + k + l}]$$
$$\times \langle O^K{}_Q \rangle\, [1 + \sigma_\pi (-1)^h \exp(i\pi l/2)]. \qquad (3)$$

The final form exploits $\langle O^K{}_Q \rangle = [\sigma_\pi\ \sigma_\theta\ (-1)^K\ \langle O^K{}_{-Q} \rangle]$ from position symmetry $m_\xi'$ that is identical in the two magnetic structures under discussion. A second element from position symmetry $2_\zeta$ requires even Q. Orthorhombic Miller indices $h$, $k$, $l$ are listed for Domains A and C in an Appendix. The time signature $\sigma_\theta = -1$ for parity-even and parity-odd magnetic absorption events. Most notably, magnetic Bragg spots enhanced by parity-even absorption events (E1 or E2) are created by electronic multipoles with odd K, i.e., $[\sigma_\theta (-1)^K = +1]$ [16]. In the present case with odd ($h + k + l$), the condition $\sigma_\theta = -1$ is enforced by body-centre anti-translation that is an explicit factor in Eq. (3) and all other electronic structure factors under investigation.

1. Domain A [3]; orthorhombic reflection vectors (0, 0, –9) & (0, 0, –11)

Explicit calculations based on Eq. (3) return null amplitudes in channels with unrotated polarization for parity-even absorption events [18]. In detail, $(\sigma'\sigma) = (\pi'\pi) = 0$ for E1-E1 and E2-E2 absorption events. The result $(\pi'\pi) = 0$ agrees with limited available experimental data [3]. The corresponding amplitudes for rotated polarization are,

$$(\pi'\sigma)_{11} = (\sigma'\pi)_{11} = (\pm)\, i\sin(\theta)\, \langle T^1{}_0 \rangle, \qquad (4)$$

with upper and lower signs for $l = -9$ and $l = -11$, respectively. A subscript 11 on the diffraction amplitude denotes an E1-E1 absorption event, and an extension of the labelling is used for E2-E2 and E1-E2 in subsequent text. Schmitt *et al.* [19] summarise some technical advantages offered by measuring intensity in the rotated channel of polarization. The two E1-E1 amplitudes do not depend on the azimuthal angle $\psi$, because Q = 0 represents complete angular isotropy. The axial dipole $\langle T^1{}_0 \rangle$ is oppositely aligned to the $a$ axis in the cubic parent structure.

An E2-E2 absorption event includes magnetic multipoles K = 1 and 3 [16, 18]. Coefficients of multipoles $\langle T^1{}_0\rangle$ and $\langle T^3{}_0\rangle$ likewise do not depend on $\psi$, while the coefficient of $\langle T^3{}_{+2}\rangle'$ contains $\cos(2\psi)$. In full, calculated magnetic E2-E2 amplitudes are $(\sigma'\sigma)_{22} = (\pi'\pi)_{22} = 0$ and,

$$(\pi'\sigma)_{22} = (\sigma'\pi)_{22} = (\pm)\, i[-\sin(3\theta)\, \langle T^1{}_0\rangle$$

$$+ \sin(\theta)\, \{(3\cos^2(\theta) - 2)\, \langle T^3{}_0\rangle - \sqrt{30}\cos^2(\theta)\cos(2\psi)\, \langle T^3{}_{+2}\rangle'\}]. \quad (5)$$

The phase convention for real ($\langle O^K{}_Q\rangle'$) and imaginary ($\langle O^K{}_Q\rangle''$) parts of a multipole is $\langle O^K{}_Q\rangle = [\langle O^K{}_Q\rangle' + i\langle O^K{}_Q\rangle'']$. At the origin of an azimuthal angle scan $\psi = 0$, and the crystal *a* axis and z axis in Fig. 1 are parallel. By design, axial amplitudes Eqs. (4) and (5) for $P_I$nna are correct for magnetic symmetry $P_I$nn2. However, Dirac amplitudes that are next in our line of study reflect the different symmetries in that their signs do not change with *l* as in Eqs. (4) and (5).

Dirac multipoles $\langle \mathbf{G}^K\rangle$ obey position symmetry $\langle G^K{}_Q\rangle = [(-1)^K \langle G^K{}_{-Q}\rangle]$ and even Q. The triangle rule for vectors shows that K = 1, 2, 3 for an E1-E2 absorption event that we study. An anapole $\langle \mathbf{G}^1\rangle$ is not permitted for $EuTi_2Al_{20}$, since $\langle G^1{}_0\rangle = 0$ by virtue of the foregoing position symmetry. As with parity-even absorption events, parity-odd amplitudes $(\sigma'\sigma)_{12} = (\pi'\pi)_{12} = 0$ [18]. Quadrupoles and an octupole diffract in the rotated channels of polarization,

$$(\pi'\sigma)_{12} = (\sigma'\pi)_{12} = i[-(3\cos^2(\theta) - 2)\, \langle G^2{}_0\rangle \quad (6)$$

$$+ 2\sqrt{(2/3)}\cos^2(\theta)\cos(2\psi)\, \{\langle G^2{}_{+2}\rangle' + 2\sqrt{2}\, \langle G^3{}_{+2}\rangle''\}].$$

Parity-even amplitudes Eqs. (4) and (5) and the parity-odd amplitude Eq. (6) are all purely imaginary, and interfere in the diffraction amplitude created by their sum. In consequence, Dirac multipoles force a difference in $|(\pi'\sigma)|^2$ between $l = -9$ and $l = -11$.

2. Domain C [3]; orthorhombic reflection vectors (5, −4, 0) & (6, −5, 0)

Diffraction is permitted with $[\sigma_\pi (-1)^h] = +1$ for $l = 0$. Specifically, axial amplitudes using multipoles $\langle \mathbf{T}^K\rangle$ are zero for odd *h*, and (5, −4, 0) is a particular case. Observed Bragg spots can be attributed to diffraction by Dirac multipoles. Wyckoff positions 8a are acentric and $\sigma_\pi = -1$ in Eq. (2).

Recall that local axes ($\xi$, $\eta$, $\zeta$) coincide with (x, y, z) in Fig. 1 in the nominal setting of the sample. The reflection vector of interest (*h*, *k*, 0) and the x axis enclose an angle $\chi$, and $\Psi^K{}_Q \rightarrow [\exp(iQ\chi)\, \Psi^K{}_Q]$ following the required alignment of the reflection vector and the x axis [18]. The crystal *a* axis is normal to the plane of scattering in Fig. 1 for $\psi = 0$. Axial amplitudes for an E1-E1 absorption event do not depend on $\chi$ because they use projections Q = 0. Results are, $(\sigma'\sigma)_{11} = 0$,

$$(\pi'\pi)_{11} = i\sin(2\theta)\cos(\psi)\, (-1)^n\, \langle T^1{}_0\rangle, \quad (7)$$

$$(\pi'\sigma)_{11} = -(\sigma'\pi) = -i\cos(\theta)\sin(\psi)\, (-1)^n\, \langle T^1{}_0\rangle, \quad (8)$$

for (2*n*, 2*m* + 1, 0).

Conversely, Dirac multipoles ($\sigma_\pi = -1$) contribute to diffraction patterns with odd *h*, i.e., (5, −4, 0). Continuing with our use of amplitudes tabulated by Scagnoli and Lovesey [18],

$$(\sigma'\sigma)_{12} = \exp(i\pi h/2) \sin(2\chi) \cos(\theta) \sin(\psi) [\langle G^2_{+2}\rangle' - \surd 2 \{1 - 3 \cos^2(\psi)\} \langle G^3_{+2}\rangle'']. \quad (9)$$

The unrotated amplitude $(\pi'\pi)_{12}$ is similar to Eq. (9). Not surprisingly, the expression for $(\pi'\sigma)_{12}$ is even more complicated with a contribution from $\langle G^2_0\rangle$ and additional contributions in $\langle G^2_{+2}\rangle'$ and $\langle G^3_{+2}\rangle''$.

**C. Magnetic symmetry $P_Inn2$** (No. 34.164, orthorhombic, polar, mm21′). Basis vectors with respect to the cubic parent structure are (½(*b* − *c*), ½(*b* + *c*), +*a*) [3]. From $2_{\xi\eta}${(½(*b* − *c*), ½(*b* + *c*), +*a*)} = {(½(*b* + *c*), ½(*b* − *c*), −*a*)}, where $2_{\xi\eta}$ is a dyad operator on a diagonal in the b-c plane, it follows that multipoles in space groups $P_Inna$ and $P_Inn2$ are related by $2_{\xi\eta}\langle O^K_Q\rangle = [\exp(i\pi Q/2)\ (-1)^K \langle O^K_{-Q}\rangle]$. Thus, isotropic multipoles $\langle O^K_0\rangle$ in the two magnetic symmetries differ in sign $(-1)^K$ alone.

Europium ions occupy Wyckoff positions 2a and 2b in $P_Inn2$, with coordinates (0, 0, 1/8) and (0, 1/2, 7/8), respectively [3]. Position symmetries for Eu ions in $P_Inna$ and $P_Inn2$ are identical. The electronic structure factor is taken to be,

$$\Psi^K_Q(P_Inn2) = [\Psi^K_Q(2a) - \Psi^K_Q(2b)] = \langle O^K_Q\rangle \exp(i\pi l/4)\ [1 + \sigma_\theta (-1)^{h+k+l}]$$
$$\times [1 - \sigma_\theta (-1)^h \exp(i\pi l/2)]. \quad (11)$$

The relative sign of the two positions in $\Psi^K_Q(P_Inn2)$ is chosen such that with $\sigma_\pi = +1$, $\sigma_\theta = -1$, magnetic symmetries $P_Inna$ and $P_Inn2$ possess identical diffraction patterns, in accord with observations [3]. Dirac amplitudes are discussed for an E1-E2 absorption event [16,18]. Actual expressions for the candidate symmetries are similar but they possess different diffraction conditions in some cases.

1. Domain A [3]; orthorhombic reflection vectors (0, 0, −9) & (0, 0, −11)

Dirac amplitudes are $(\sigma'\sigma)_{12} = (\pi'\pi)_{12} = 0$, and Eq. (6) holds for $(\pi'\sigma)_{12} = (\sigma'\pi)_{12}$ after a simple change in overall sign.

2. Domain C [3]; orthorhombic reflection vectors (5, −4, 0) & (6, −5, 0)

Dirac multipoles for an E1-E2 absorption event are permitted for even *h*, and (5, −4, 0) is a forbidden reflection. The unrotated amplitude $(\sigma'\sigma)_{12}$ is the same as Eq. (9) after removing the phase factor exp(*iπh*/2).

## III. DISCUSSION AND CONCLUSIONS

We have used symmetry informed electronic structure factors to unveil properties of $EuTi_2Al_{20}$ encapsulated in candidate magnetic symmetries [3]. The room-temperature parent structure is the cubic diamond structure with Eu ions in asymmetric Wyckoff positions (No. 227). It is epitomized by Bragg spots that are forbidden by spherically symmetric ions, i.e., intensities are due to $sp^3$ hybridized diamond ions [4,5]. Long-range magnetic order in $EuTi_2Al_{20}$ appears below a temperature ≈ 3.3 K. Two candidates, one centrosymmetric (No.

52.320 [12]) and one polar (No. 34.164), have been inferred from solid experimental evidence obtained from the diffraction of neutrons and x-rays [3]. Their properties are deeply hidden by time reversal symmetry 1′ in their magnetic crystal classes, and concomitant body-centre anti-translation. We show that magnetic entities with discrete symmetries that match those of the Dirac monopole tell the two symmetries apart [8, 16,18]. To this end, we contrast calculated resonant x-ray diffraction patterns for the two candidate magnetic symmetries. Anapoles (Dirac dipoles) are not allowed in the candidates. Notably, anapoles have been directly observed in magnetic neutron diffraction by magnetic symmetry mm′m′ (position symmetry mm′2′) derived from the cubic diamond structure [11]. Our x-ray patterns for $EuTi_2Al_{20}$ include an azimuthal angle and a full polarization analysis.

**Acknowledgements** Dr M. Kawamata supplied basis vectors for the two orthorhombic candidate symmetries. Insightful correspondence with Dr R. D. Johnson and Professor S. P. Collins.

## APPENDIX: ORTHORHOMBIC REFLECTION VECTORS

Integer Miller indices (*h*, *k*, *l*) for orthorhombic magnetic structures are permitted for odd (*h* + *k* + *l*), i.e., body-centre anti-translation that is explicit in electronic structure factors Eq. (3) and Eq. (11). For $P_Inna$ or $P_Inn2$ symmetry, one requires propagation vectors $k_1$ = (1, 0, 0), $k_2$ = (0, 1, 0) and $k_3$ = (0, 0, 1). Unit cell transformations for these domains constructed from orthogonal vectors are {(0, 1/2, 1/2), (0, 1/2,−1/2), (−1, 0, 0)}, {(1/2, 0, 1/2), (1/2, 0,−1/2), (0, 1, 0)} and {(1/2, 1/2, 0), (−1/2, 1/2, 0), (0, 0, 1)}, respectively. A general cubic reflection ($H_o$, $K_o$, $L_o$) with integer Miller indices appears in Section II.A. The cubic reflection (8, 1, 0) can be presented as [(8, 0, 0) + $k_2$]. This implies use of $k_2$ and the associated transformation takes the (8, 1, 0) cubic reflection into orthorhombic (4, 4, 1). Likewise, cubic (10, 1, 0) is transformed into orthorhombic (5, 5, 1) in the same $k_2$ domain. Cubic (9, 1, 0) can be presented as [(9, 1,−1) + $k_3$], implying the domain with $k_3$ propagation vector. The associated transformation takes cubic (9, 1, 0) into orthorhombic (5,−4, 0). Cubic (11, 1, 0) is [(11, 1,−1) + $k_3$], and the associated transformation takes it orthorhombic (6,−5, 0). Cubic reflections (9, 0, 0) and (11, 0, 0) can be presented as [(8, 0, 0) + $k_1$] and [(10, 0, 0) + $k_1$], respectively. The associated transformation takes them to (0, 0, −9) and (0, 0, −11). Associated transformations are identical for orthorhombic symmetries $P_Inna$ and $P_Inn2$. In consequence, orthorhombic Miller indices are identical for both magnetic symmetries.

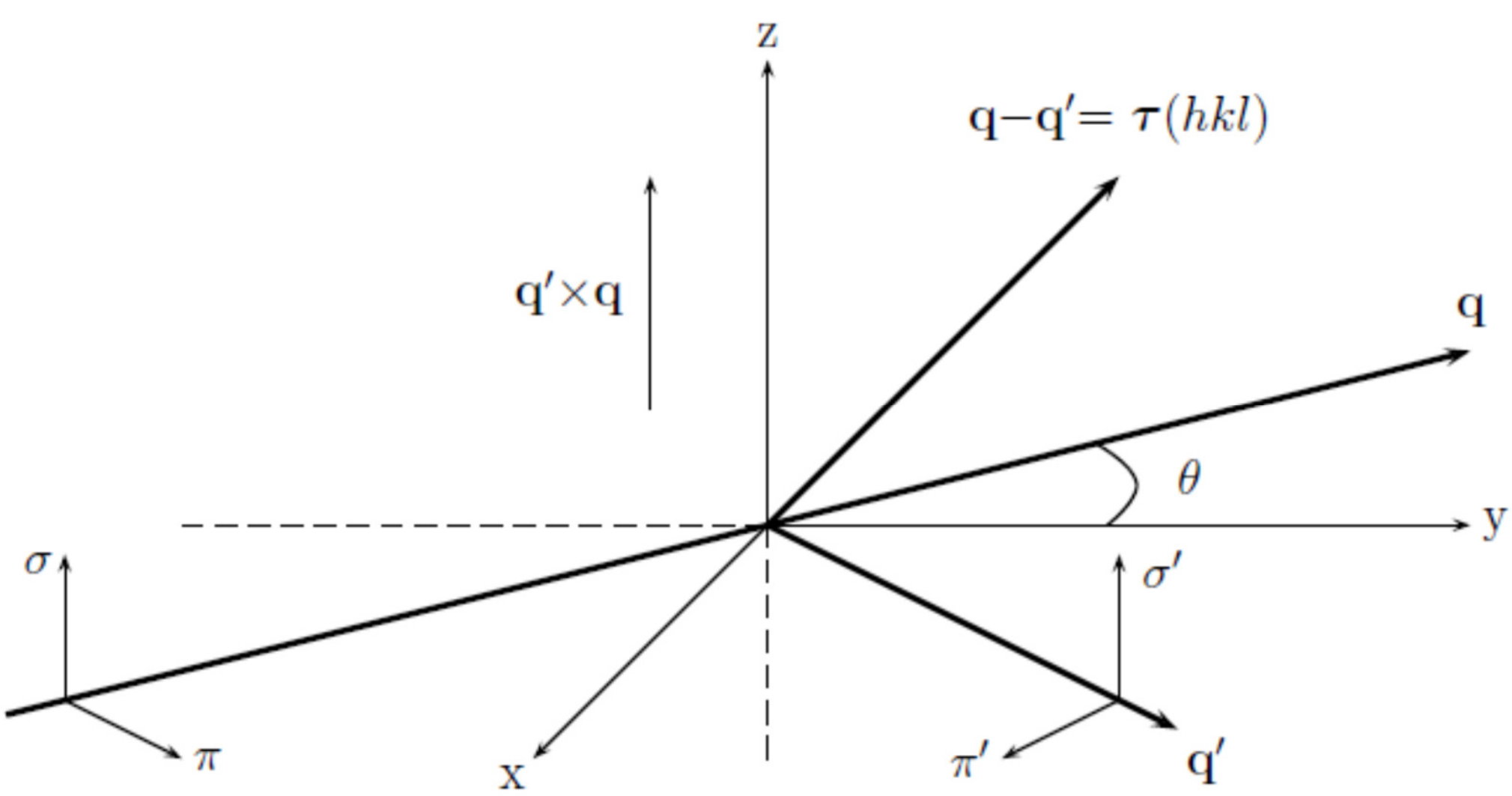

**FIG. 1**. Primary (σ, π) and secondary (σ′, π′) states of x-ray polarization; **σ** = **σ′** = (0, 0, 1), **π** = (cos(θ), sin(θ), 0), **π′** = (cos(θ), − sin(θ), 0). Corresponding wavevectors **q** and **q′** subtend an angle 2θ. The Bragg condition for diffraction is met when **q** − **q′** coincides with a vector **τ**(*h*, *k*, *l*) of the reciprocal lattice. The depicted Cartesian (x, y, z) coincide with local axes (ξ, η, ζ) in the nominal setting of the crystal.